# Bhirkuti's Test of Bias Acceptance (BTBA): Examining Its Performance in Psychometric Simulations

Aneel Bhusal[1], Todd Little[1,2]
[1]Department of Educational Psychology, Leadership, and Counseling, Texas Tech University
[2]Optentia Research Focus Area, North-West University, Vanderbijlpark, South Africa
Correspondence concerning this article should be addressed to
Aneel Bhusal, Email: aneelbhusal@gmail.com

**Abstract**

We introduce *Bhirkuti's Test of Bias Acceptance (BTBA)*, a standardized framework for evaluating estimator bias in Monte Carlo simulation studies. BTBA uses a simulation-specific standardized score (Z*) and a decision matrix to assess bias acceptability based on the mean and variance of Z* distributions. Under ideal conditions, Z* values should approximate a standard normal distribution (Z-distribution) with a mean near zero and variance near one in the context of simulation research. Systematic deviations from these patterns such as shifted means or inflated variances indicate bias or estimator instability in simulation-based research. BTBA visualizes these patterns using ridgeline density plots, which reveal distributional features such as central tendency, spread, skewness, and outliers. Demonstrated in a latent growth modeling context, BTBA offers a reproducible and interpretable method for diagnosing bias across varying simulation conditions. By addressing key limitations of traditional relative bias (RB) metrics, BTBA provides a theoretically grounded, distribution-aware, transparent, and replicable alternative for evaluating estimator quality, particularly in psychometric modeling, structural equation modeling, and missing data research. Through this framework, we aim to enhance methodological decision-making by integrating statistical reasoning with comprehensive visualization techniques.

## Introduction

**Overview of Relative Bias (RB) in Simulation-Based Bias Evaluation**

Relative Bias (RB) is a commonly used metric in simulation-based methodological research to quantify the average proportional deviation of an estimated parameter from its true population value. It is calculated by subtracting the true parameter ($\theta$) from the estimated parameters ($\hat{\theta}$) and dividing the difference by the true parameter (see Equation 1). The result is typically expressed as a percentage. This measure provides an interpretable way to evaluate estimator performance by indicating the degree and direction of systematic overestimation or underestimation. Relative bias (RB) has been widely applied in simulation studies of structural equation models, latent growth models, and missing data methodologies, where comparing bias across various estimators or study conditions is essential for methodological assessment (Enders & Bandalos, 2001; Little & Rubin, 2019; McArdle & Epstein, 1987; Rubin, 1987).

$$RB_\theta = \frac{\hat{\theta} - \theta}{\theta} \tag{1}$$

**Limitations of Relative Bias (RB)**

Relative Bias (RB) is subject to several limitations that can compromise its interpretability in simulation-based estimator evaluations. These limitations include:

*Scale sensitivity:* RB is highly sensitive to the magnitude of the population parameter. Because it divides the estimation error by the true parameter value, the same absolute bias yields very different RB values depending on that magnitude. For instance, an absolute deviation of 0.005 results in an RB of 5% when the population parameter is 0.1, but only 0.9% when the parameter is 0.55. This scaling property inflates bias for smaller parameters and deflates it for larger ones, potentially misrepresenting the practical significance of estimation error (Garnier-Villarreal et al. ,2014; Lang et al., 2020). For example, in a latent growth model estimating slope correlations, a consistent bias of 0.005 would be flagged as highly problematic if the population correlation is 0.1 yet dismissed as insignificant if the correlation is 0.55 even though the estimation error is identical.

*Relative Bias Overstates or Understates Bias Depending on Parameter Size:* Relative Bias (RB) can be misleading because it reacts differently depending on the size of the true population parameter. When the true value is small, even tiny bias in estimation can result in large RB values. For example, if the true value is 0.02 and the estimate is off by just 0.004, the RB becomes 20%, which may wrongly suggest that the estimator performs poorly. This exaggeration is common in models with small effect sizes. On the other hand, when the true parameter is large, RB tends to hide meaningful bias. For instance, if the true value is 0.8 and the estimate is off by 0.04, the RB is just 5%, which might seem acceptable even though the absolute error is notable. This pattern is especially problematic in simulation studies that compare results across parameters of different magnitudes, because RB inflates bias for small values and deflates it for large ones creating a distorted view of estimator performance (Rhemtulla et al., 2014).

*Instability near zero:* RB becomes mathematically unstable or undefined when the population parameter approaches zero. In such cases, even minute deviations can produce extremely large or infinite RB values. For example, if a parameter such as an intercept or residual variance has a population value near 0.001, an estimate of 0.005 would yield an RB of 400%, despite the deviation being practically negligible. This instability renders RB unusable for parameters near zero, which are common in models involving baseline effects, residual variances, or small mediating paths (Rhemtulla et al., 2014).

*Neglect of Estimate Variability:* Relative Bias (RB) captures only the average percentage deviation of estimates from the true value, but it ignores how much estimates vary across replications (see Equation 1). This neglect can create a false impression of stability. For instance, an estimator might show a small average RB say, 2% but if its estimates range widely from -0.20 to +0.25, some are still far from the true value. In this case, RB hides the fact that the estimator is inconsistent. Consider a simulation where the true slope is 0.30, and across replications, the estimates vary between 0.10 and 0.50. Even if the average estimate is close to 0.306 (RB ≈ 2%), the large spread indicates high uncertainty and low reliability. This limitation is especially critical in small-sample simulations or complex models that cause estimates to fluctuate widely around the population value.

**Bhirkuti's Test of Bias Acceptance (BTBA): A Standardized and Visual Framework for Evaluating Simulation Bias**

*Z\* as a Standardized Metric for Bias Assessment*

To address the limitations of traditional bias metrics in simulation research, this study uses Z* as a robust, standardized measure of bias centered on the known true parameter value (see Equation 2). Unlike conventional z-scores, which standardize around sample means and standard deviations, Z* is specifically designed for simulation contexts where the true population value (θ) is known and fixed. Z* quantifies the deviation of each estimate ($\hat{\theta}$) from true population value (θ), scaled by the root mean squared error (RMSE) across all replications. RMSE shows how much estimates typically differ from the true value, making it easier to compare bias across different parameters and conditions, no matter their scale.

$$Z^* = \frac{Bias}{RMSEA} = \frac{Bias}{\sqrt{Bias^2 + Variance}} = \frac{\hat{\theta} - \theta}{\sqrt{\frac{1}{R}\sum_{r=1}^{R}(\hat{\theta} - \theta)^2}} \quad (2)$$

Where:

- $\hat{\theta}$: Estimated value
- $\theta$: True population value
- R: Total number of converged replications

By anchoring each estimate to the true population value and adjusting for overall estimation variability, Z* offers an intuitive bias metric aligned with standard inferential logic. Under ideal conditions, the distribution of Z* is expected to have a mean near zero and variance near one, mirroring the properties of a standard normal distribution (Bollen, 1989; Mooney, 1997). This property makes Z* particularly useful for simulation diagnostics, where it supports meaningful cross-condition comparisons of estimator performance. Given the large number of replications (5,000 per condition), the distribution of Z* naturally stabilizes and approximates normality under the central limit theorem (Bollen, 1989; Mooney, 1997). As such, BTBA does not require strict normality assumptions for individual estimates; rather, it evaluates the collective behavior of Z* using its mean and variance as summary diagnostics.

*BTBA-Inspired Decision Matrix for Z\* as a Standardized Metric Bias Acceptability*

To operationalize Z* as a diagnostic tool, this study proposes a decision matrix grounded in Bhirkuti's Test of Bias Acceptance (BTBA). The matrix defines bias acceptability zones based on the mean and variance of Z* across replications, leveraging its standard normal-like distributional behavior (Bollen, 1989; Mooney, 1997). When estimator performance is ideal, Z* values should center around zero with unit variance, and deviations from this benchmark suggest potential bias or instability.

The BTBA decision matrix provides five interpretive zones:

| Mean of $Z^*$ | Variance of $Z^*$ | Interpretation | Bias Acceptability | BTBA Verdict |
|---|---|---|---|---|
| -0.10 to 0.10 | 0.90 to 1 | Near-zero bias and stable variance | Accept Bias | Accept |
| -0.20 to -0.10 or 0.20 to 0.10 | 0.90 to 1 | Mild bias, within acceptable error margin | Accept with caution | Accept |
| -0.30 to -0.20 or 0.20 to 0.30 | 0.90 to 1 | Moderate Bias; may be tolerable depending on the research | Research dependent | Research dependent |
| Beyond $\pm 0.30$ | < 0.90 | Meaningful Bias | Reject Bias | Reject |

This structured matrix allows researchers to interpret $Z^*$ distributions in the context of simulation-based estimator performance. It identifies when bias is negligible (within ±0.10), tolerable (up to ±0.30), or practically concerning (beyond ±0.30), while also flagging estimation instability when the variance of $Z^*$ falls outside the expected 0.90–1.0 range. This dual-axis decision framework brings both statistical rigor and practical interpretability to simulation diagnostics. $Z^*$ variance greater than one is rare in properly designed simulations research. Like the convention of accepting absolute relative bias below 10%, BTBA thresholds are based on the expected normality of $Z^*$ under the Central Limit Theorem (Bollen, 1989; Mooney, 1997). Further research is needed to examine their applicability across diverse contexts.

**Comparison of Z* and Traditional Z-Scores in Bias Interpretation**

While $Z^*$ shares mathematical similarities with the traditional z-score, it serves a fundamentally different purpose tailored for simulation-based evaluation. The standard z-score is designed for inferential contexts where the population mean is unknown, and the focus is on an individual observation's deviation from the sample mean, scaled by the sample standard deviation. In contrast, $Z^*$ is constructed specifically for simulation scenarios where the true population parameter (θ) is known, enabling direct and meaningful bias evaluation (see Equation 2).

Rather than centering on an empirical mean, $Z^*$ centers each estimate on the known true value θ and scales it using the RMSE, which captures the typical estimation error observed across replications. This design allows $Z^*$ to quantify how much each estimate deviates from truth relative to expected simulation variability, effectively standardizing bias, not raw observations (see Equation 2). Under ideal estimator behavior, the distribution of $Z^*$ should approximate a standard normal distribution, with a mean near zero and variance near one allowing researchers to use familiar inferential benchmarks to determine when bias becomes problematic (Bollen, 1989; Mooney, 1997).

This property makes $Z^*$ not only theoretically grounded but also practically interpretable, as it allows bias to be judged within a framework already familiar to researchers trained in inferential statistics. By translating raw estimation errors into standardized bias scores, Z* bridges the gap between traditional hypothesis testing logic and simulation diagnostics.

**Visual Distribution of Parameter Estimates and Z* Values: Use of Ridgeline Plots**

To complement the standardized bias diagnostics, this study employs ridgeline plots to visualize the full empirical distribution of parameter estimates across experimental conditions. These plots provide a distributional perspective that extends beyond traditional point-based metrics such as Relative Bias (RB) and Absolute Relative Bias (ARB), which can obscure variability and distributional characteristics. By displaying the shape, spread, and central tendencies of estimates generated from Monte Carlo simulations, ridgeline plots enable a more comprehensive evaluation of estimator behavior.

Ridgeline plots are particularly effective in identifying key features of the simulation output, including systematic trends, outliers, distributional skew, multimodality, and estimation stability across replications (Wilke, 2021). These visual cues reveal not only the presence or absence of bias but also the consistency and reliability of estimators under varying model assumptions, sample sizes, and missing data mechanisms (Bhusal, 2024). This visual context is essential for distinguishing between random error and systematic estimation shifts that may indicate model misfit or design inefficiency.

To further enhance interpretability, each ridgeline plot can incorporate relevant descriptive statistics, including the mean, median, mode, and variance of the parameter estimates (Bhusal, 2024). These statistics provide concrete reference points for assessing the magnitude and direction of deviations from the true value. Moreover, plotting Z* values alongside the distributions helps contextualize bias in standardized terms, facilitating direct comparisons to the BTBA decision matrix thresholds.

**Bhirkuti's Test of Bias Acceptance (BTBA): Integrative Framework for Robust Simulation Evaluation**

Bhirkuti's Test of Bias Acceptance (BTBA) provides a comprehensive and scalable framework for evaluating bias in simulation studies by integrating three complementary components: a standardized bias metric (Z*), a theoretically grounded decision matrix for bias classification, and a visual distributional assessment through ridgeline plots. This integrative approach addresses key limitations of traditional bias diagnostics by uniting standardized quantification, statistical interpretability, and visual transparency within a single evaluative system. The framework is particularly well-suited for complex modeling scenarios such as those involving planned missing data designs, small sample sizes, or high-dimensional parameter structures where traditional bias measures often fall short. Through its structured logic and emphasis on distributional behavior, BTBA enhances the rigor, robustness, and clarity of simulation-based estimator evaluation.

# Simulation Study

To evaluate the performance of Bhirkuti's Test of Bias Acceptance (BTBA), this study employs a Monte Carlo simulation (Carsey & Harden, 2013) using a Latent Growth Curve Model (LGM) (McArdle et al., 1987). The simulation examines estimator bias under planned missing data with FIML compared to complete data, and assesses whether BTBA provides a clearer, distribution-aware evaluation (Graham et al., 2001; Rhemtulla et al., 2014).

**Latent Growth Curve Model (LGM)**

The simulated data are based on a bivariate Latent Growth Model (LGM) representing two developmental constructs: Bullying (B) and Homophobic Teasing (H). These latent constructs capture changes in behavior over five equally spaced time points (Little, 2024). Each construct is measured using three observed indicators per wave, offering sufficient measurement precision while maintaining model parsimony. The data generation process aligns with methods described in Rhemtulla et al. (2014) and Little (2024), ensuring compatibility with established simulation protocols. Population values for growth factor means, variances, and covariances were derived from prior empirical work in longitudinal developmental psychology (Rhemtulla et al., 2014).

**Figure 1**

*Proposed Growth Curve Model for Simulation*

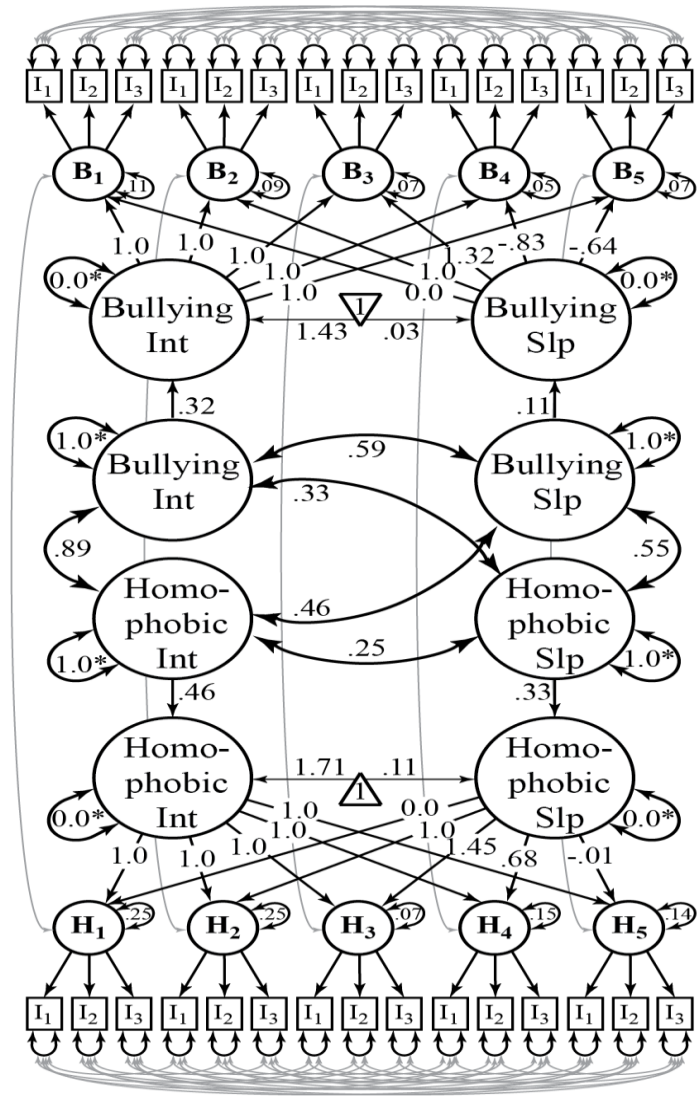

**Simple Wave Missing Design (SWMD-6)**

To systematically assess how planned missing data affects estimator performance, this study adopts the Simple Wave Missing Design (SWMD-6), a structured approach to missingness introduced by Graham et al. (2001). SWMD-6 is commonly used in longitudinal studies to enhance data collection efficiency without compromising statistical power or parameter estimation. The design involves six participant groups, each assigned a specific pattern of missingness across five time points. One group (Group 1) provides complete data across all occasions, while the remaining five groups (Groups 2–6) follow varied, pre-defined missing data schedules.

**Table 1**

*Simple Wave Missing Design with Six Groups (SWMD-6)*

| Random Group | Occasion Of Measurement | | | | |
|---|---|---|---|---|---|
| | Time 1 | Time 2 | Time 3 | Time 4 | Time 5 |
| 1 | √ | √ | √ | √ | √ |
| 2 | √ | √ | √ | √ | X |
| 3 | √ | √ | √ | X | √ |
| 4 | √ | √ | X | √ | √ |
| 5 | √ | X | √ | √ | √ |
| 6 | X | √ | √ | √ | √ |

*Note:* Within each time occasion, √ = data present, and X = data missing; (Graham et al., 2001; Little, 2024)

This structured approach ensures that each time point is adequately represented across the sample, minimizing information loss while reducing participant burden. The design supports robust estimation by preserving temporal coverage despite planned gaps in data. Notably, SWMD-6 is compatible with Full-Information Maximum Likelihood (FIML) estimation, which can produce unbiased estimates using all available data (Rhemtulla et al., 2014). Its stability under FIML makes it a practical and effective choice in applied research. Within this framework, the study evaluates Bhirkuti's Test of Bias Acceptance (BTBA) across multiple planned missingness conditions, allowing for a rigorous assessment of the method's reliability and applicability in bias evaluation within simulation research.

**Full-Information Maximum Likelihood (FIML) Estimation**

Full-Information Maximum Likelihood (FIML) is a widely adopted method for addressing missing data, offering direct parameter estimation by leveraging all available data without requiring explicit imputation. In contrast to multiple imputations (MI), which creates complete datasets by replacing missing values with plausible estimates, FIML constructs a likelihood function for each observed response pattern and estimates model parameters by maximizing the combined likelihood. This approach enhances estimation efficiency while preserving statistical integrity, making it especially suitable for structural equation modeling and longitudinal research (Dempster et al., 1977; Enders et al., 2001; Enders, 2022). FIML yields unbiased estimates under missing completely at random (MCAR) and missing at random (MAR) conditions. In particular, the MAR assumption permits valid inference by allowing the missingness mechanism to be fully accounted for through observed variables (Enders, 2022).

## Simulation Design and Conditions

To rigorously evaluate the performance of Bhirkuti's Test of Bias Acceptance (BTBA), a series of Monte Carlo simulations were conducted under systematically varied conditions. The simulation design manipulated latent slope correlations, sample sizes, and data structures using the Simple Wave Missing Design (SWMD-6) and a complete data condition to examine estimator performance across a range of analytic contexts. Latent slope correlations ($\rho_{s1,s2}$) were specified at three levels: 0.1, 0.3, and 0.55, reflecting weak to moderate associations between latent growth trajectories. Sample sizes for each of the six groups varied from small (n = 40, 60, 80, 100) to moderate (n = 300, 500) and large-scale conditions (n = 800, 1000), enabling a comprehensive assessment of estimator performance consistent with previous study (Rhemtulla et al., 2014). Full-Information Maximum Likelihood (FIML) was used to estimate model parameters under the planned missingness structure defined by the SWMD-6 design. Results obtained from this condition were compared to those derived from a complete data scenario to evaluate the impact of missingness. Each simulation condition was replicated 5,000 times to ensure stable and accurate estimation of performance metrics.

Estimator performance was evaluated using a set of criteria aligned with the Bhirkuti's Test of Bias Acceptance (BTBA) framework. First, Relative Bias (RB) was calculated to assess systematic deviation between estimated and true population values. This measure served as a baseline indicator of bias magnitude, consistent with established practices in simulation research (Muthén et al., 1987; Garnier-Villarreal et al., 2014; Lang et al., 2020; McArdle & Epstein, 1987; Moore et al., 2020). Second, graphical diagnostics were employed to visually inspect the distributional characteristics of parameter estimates across conditions. Boxplots and ridgeline plots were used to capture bias direction, magnitude, and variability, offering an intuitive understanding of estimation behavior under different simulation scenarios. Third, the core BTBA metric, $Z^*$, was computed by standardizing each parameter estimate relative to its root mean squared error (RMSE). Under ideal conditions, $Z^*$ is expected to approximate a standard normal distribution with a mean of zero and variance of one. This standardization procedure enables scale-free comparisons of bias and facilitates interpretation across varying parameter contexts. Finally, the BTBA decision matrix was applied to the distribution of $Z^*$ values. By jointly evaluating the mean and variance of the $Z^*$ distribution, the framework classifies bias into corresponding categories of acceptability: accept, accept with caution, research dependent, or reject. This classification provided a structured approach for interpreting the practical implications of bias in simulation-based estimation. All simulations and analyses were conducted in the R statistical environment using the *lavaan*, *ggplot2*, and *ggridges* packages. This comprehensive framework allowed for a systematic and interpretable evaluation of bias.

## Results

**Evaluation of Relative Bias (RB) and Limitations of Absolute Relative Bias (ARB)**

Relative Bias (RB) is a commonly used metric in simulation studies to evaluate the proportional deviation of estimated parameters from their true population values. It reflects both the direction and the magnitude of systematic bias by comparing the average estimate across replications to the population value. In many methodological studies, RB is also converted into Absolute Relative Bias (ARB) to isolate the magnitude of bias, independent of direction. ARB is calculated as the absolute value of RB and is usually expressed as a percentage. While RB provides important information about whether estimates tend to overestimate or underestimate the true value, ARB allows for a standardized summary of bias magnitude. This feature makes it useful for comparing estimator performance across parameters, especially when directionality is not of primary concern. In this study, ARB is included as a baseline descriptive measure of estimator behavior. Its interpretation, however, is limited by its sensitivity to the scale of the population parameter. As illustrated in Figure 2, even small absolute differences can result in inflated ARB values when the true parameter is small. For example, an absolute bias of 0.005 corresponds to ARB values of 5.0 percent, 1.6 percent, and 0.9 percent when the true values are 0.1, 0.3, and 0.55, respectively. This inflation for small parameters and compression for larger ones reduces the interpretability of ARB in models that include parameters with different scales. Across all sample sizes and slope correlation levels, Absolute Relative Bias (ARB) rarely fell below the commonly used 10 percent threshold for negligible bias (Muthén et al., 1987; Garnier-Villarreal et al., 2014), even when the absolute bias was small. This pattern indicates that ARB may exaggerate the severity of bias when population parameters are small, leading to inflated evaluations of estimator inaccuracy. Conversely, when population parameters are large, ARB can mask substantial absolute deviations by yielding deceptively low percentage values. Together, these tendencies suggest that ARB may both overstate and understate the seriousness of bias depending on the scale of the parameter, potentially misrepresenting estimator performance in simulation studies. These findings highlight the need for scale-invariant, bias-sensitive approaches that more accurately reflect true estimation behavior across diverse conditions.

**Figure 2**

*Absolute Relative Bias: Slope-Slope Correlation*

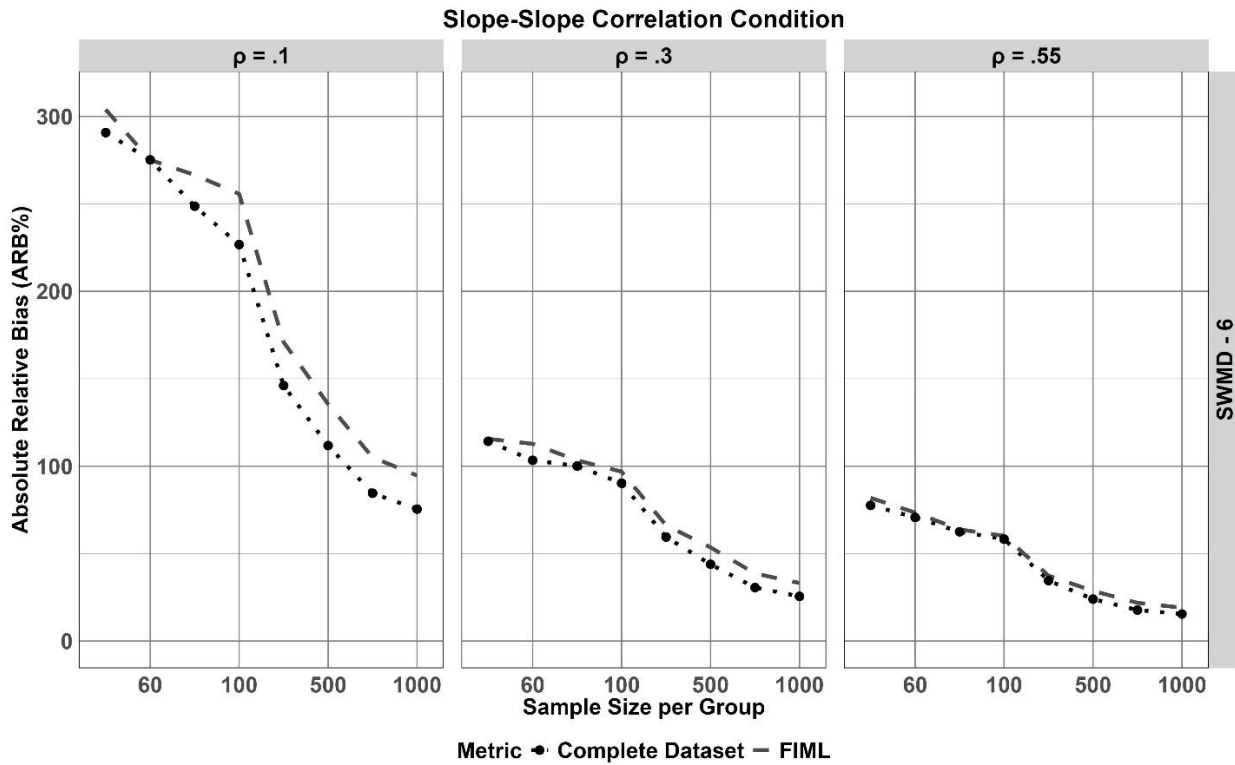

**BTBA Framework for Bias Evaluation**
    **Tool 1: Visual Diagnostics – Boxplots and Ridgeline Plots**

    Visual inspection of estimator behavior was conducted using boxplots and ridgeline plots. Ridgeline plots display the full empirical distribution of slope–slope correlation estimates across 5,000 replications for each condition, stratified by sample size, data condition (Complete vs. FIML), and population correlation level ($\rho$ = 0.1, 0.3, 0.55). Each plot includes annotations for the mean (M) and variance (V) of the distribution. As sample size increases, the distributions become more sharply peaked around the true population value, indicating improved estimator precision. This pattern is especially apparent in the complete data condition, where smaller samples yield wide, dispersed distributions and larger samples result in tighter, symmetric ones. Under the FIML condition, similar trends are observed, though distributions are generally wider in smaller samples, reflecting the additional impact of planned missingness. Vertical reference lines denote the true parameter value, while dots and horizontal bars represent the distribution mean and ±1 standard deviation, respectively. These plots offer a visual foundation for evaluating estimator stability and bias across all 24 simulated conditions.

**Figure 3**
*Ridgeline Plot of Estimate*

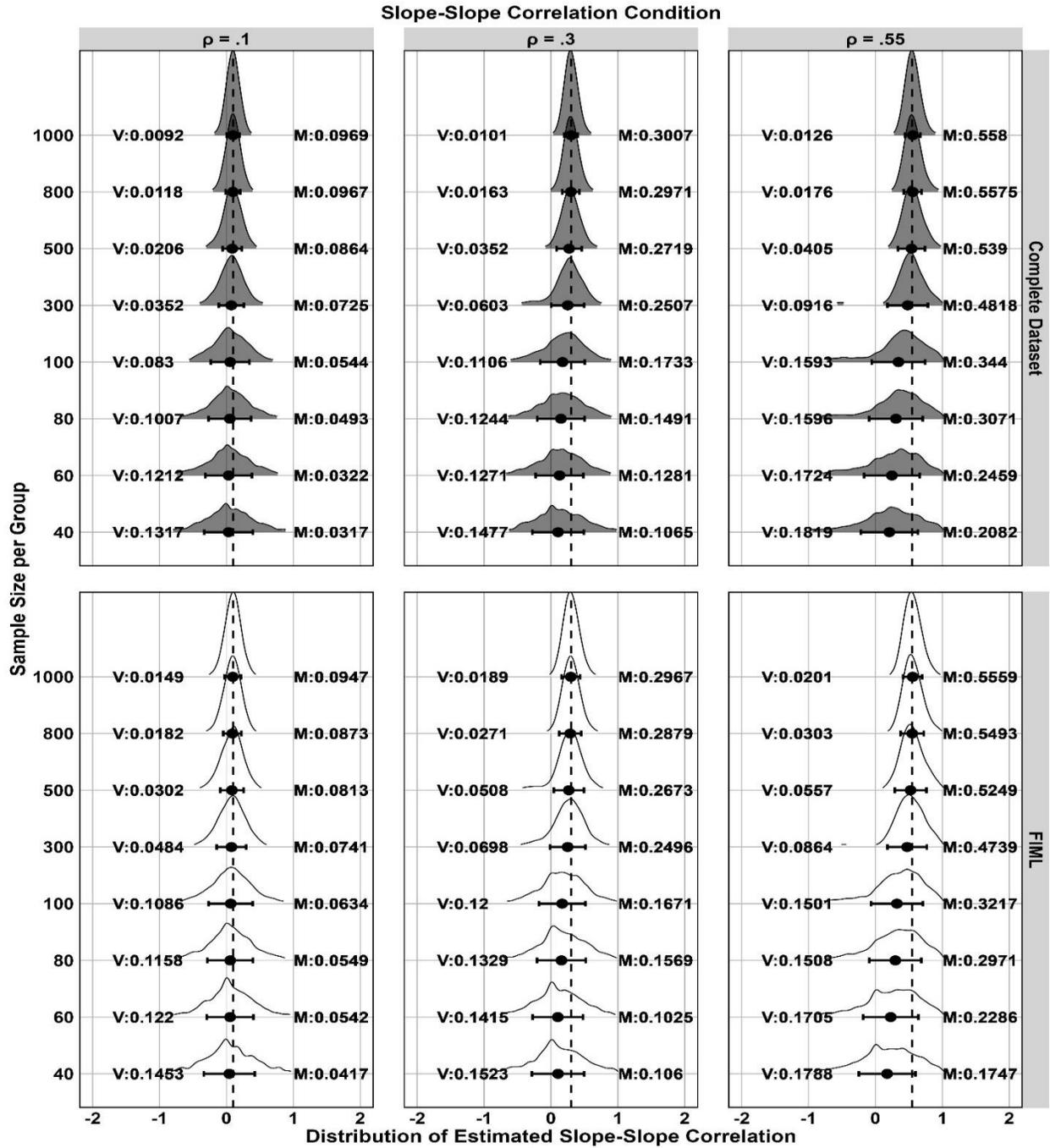

*Note:* Dots are the distribution means and bars show ±1 SD from the mean, The vertical line represents the population value. Distribution of estimate for 24 conditions. "V:" variance of the estimate produced and "M:" mean of the estimate produced.

**Figure 4**

*Box-plot along with rigid-line distribution curve for Slope-Slope Correlation*

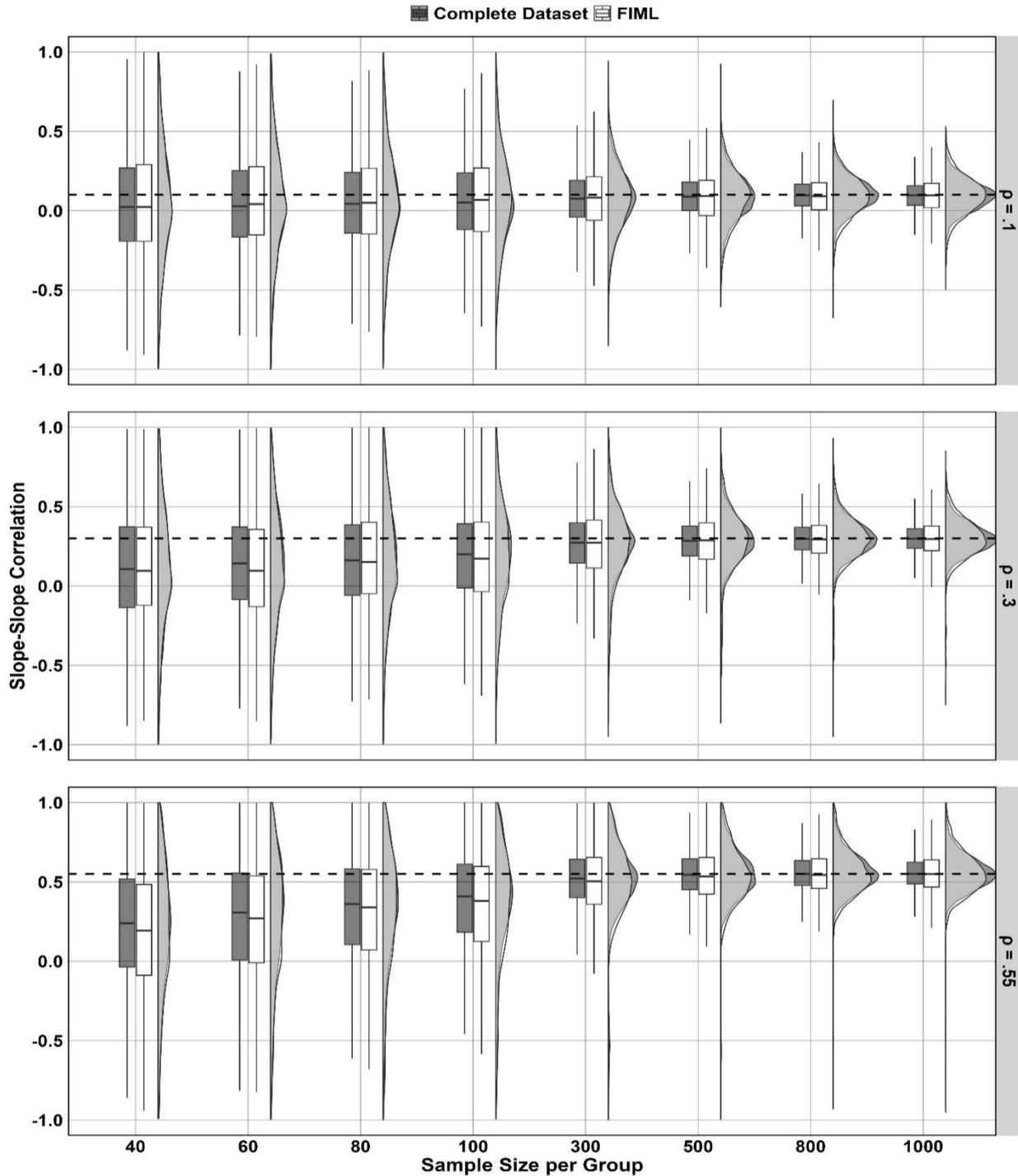

*Note:* Horizontal dashed line represents the true population value of the slope-slope correlation. Boxplots represent the distribution of estimated slope-slope correlations, with the central line indicating the median and the boxes capturing the interquartile range. Rigid-line density curves illustrate the shape of the estimate distributions.

**Tool 2: Standardized Bias (Z)***

Z* is the core diagnostic metric in the BTBA framework. It is computed by standardizing each parameter estimate relative to the root mean squared error (RMSE) across replications. Under ideal conditions, Z* values should approximate a standard normal distribution, with a mean of zero and a variance of one. The accompanying ridgeline plots illustrate the distribution of Z* values across sample sizes and population correlations, separately for complete and FIML data. For each condition, the mean (M) and variance (V) of the Z* distribution are labeled, providing the necessary inputs for BTBA classification. According to the framework, bias is considered negligible when the Z* mean falls within ±0.10 and the variance remains between 0.90 and 1.10. As shown in the plots, large sample sizes produce distributions that are more tightly centered with variances near one, particularly for $\rho = 0.1$ and $\rho = 0.3$. In contrast, smaller samples yield flatter, more variable distributions. These deviations from the ideal (e.g., means beyond ±0.10 or variances below 0.90) correspond to BTBA classifications of moderate or concerning bias.

**Figure 5**
*Ridgeline Plot of Z\**

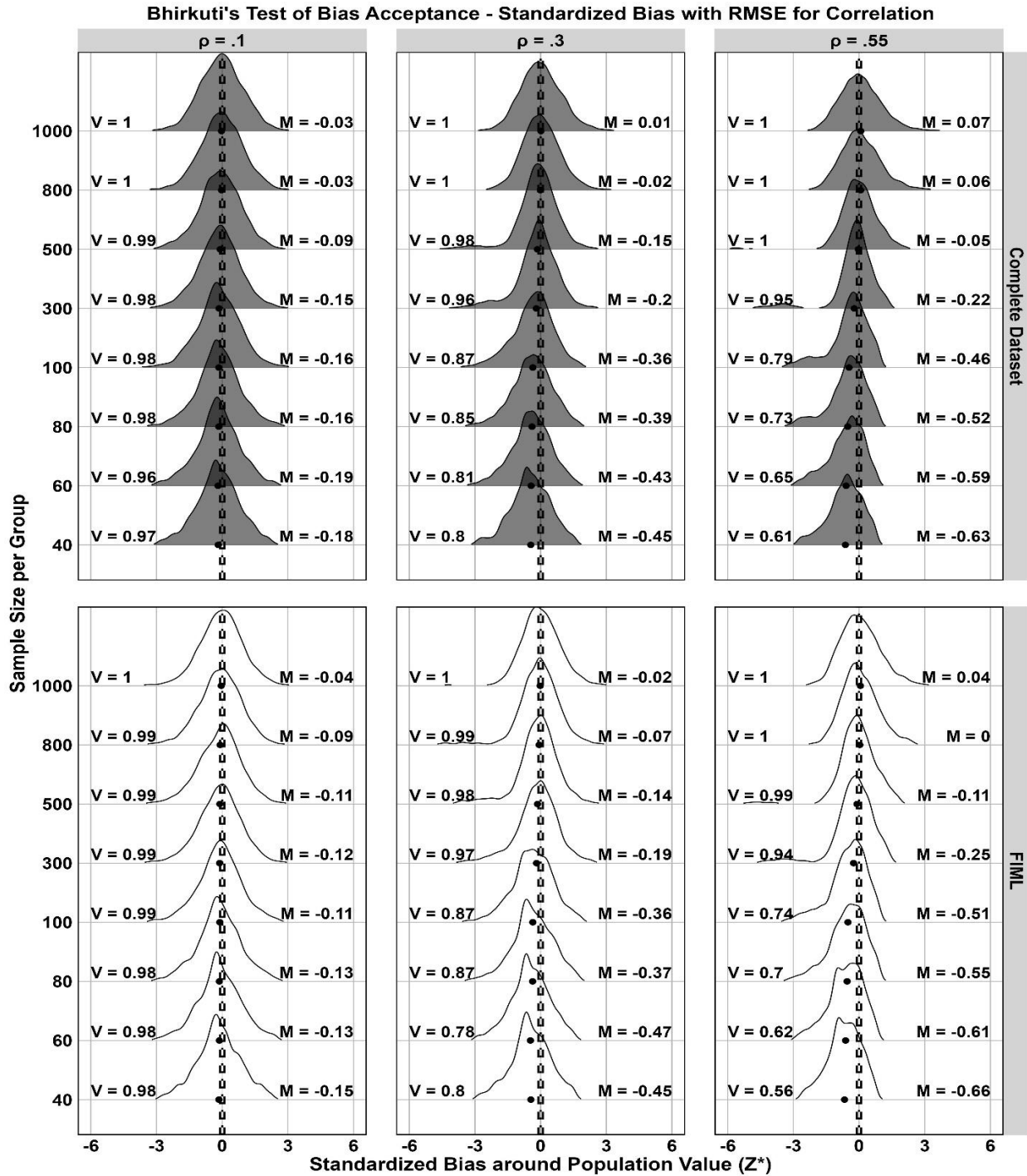

*Note:* Dots are the distribution means. The vertical black dotted line represents the idea bias line after transformation at Z\*=0. Two vertical dashed lines represent acceptable zone under BTBA at $\pm 0.1\ Z^*$. Distribution of estimate for 24 conditions. "V:" variance of the estimate produced and "M:" mean of the estimate produced.

**Tool 3: BTBA Decision Matrix**

The BTBA decision matrix integrates the mean and variance of Z* to systematically classify bias levels. The matrix identifies four zones: accept, accept with caution, research dependent, or reject based on how far Z* distributions deviate from the standard normal expectation. This structured approach enables a more interpretable and rigorous assessment of estimator quality. As depicted in the visual outputs, large samples (n ≥ 800) generally meet BTBA's acceptability thresholds under both complete and FIML conditions. However, smaller samples (n ≤ 100), particularly under FIML, often show substantial deviations. Z* means in these conditions can exceed ±0.30, and variances may fall below 0.90, signaling potential estimator instability or severe bias. The matrix provides a transparent and replicable mechanism for determining the acceptability of bias in simulation results.

Table 2 Bhirkuti's Test of Bias Acceptance (BTBA) - Verdict

| Correlation Condition / Sample Size Per Group | Complete Dataset (ρ = .1) | Complete Dataset (ρ = .3) | Complete Dataset (ρ = .55) | FIML (ρ = .1) | FIML (ρ = .3) | FIML (ρ = .55) |
|---|---|---|---|---|---|---|
| 1000 | Accept | Accept | Accept | Accept | Accept | Accept |
| 800 | Accept | Accept | Accept | Accept | Accept | Accept |
| 500 | Accept | Accept with caution | Accept | Accept with caution | Accept with caution | Accept with caution |
| 300 | Accept with caution | Accept with caution | Research dependent | Accept with caution | Accept with caution | Research dependent |
| 100 | Accept with caution | Reject | Reject | Accept with caution | Reject | Reject |
| 80 | Accept with caution | Reject | Reject | Accept with caution | Reject | Reject |
| 60 | Accept with caution | Reject | Reject | Accept with caution | Reject | Reject |
| 40 | Accept with caution | Reject | Reject | Accept with caution | Reject | Reject |

**Conclusion and Limitations**

Bhirkuti's Test of Bias Acceptance (BTBA) offers a theoretically grounded and practically interpretable framework for evaluating bias in simulation-based methodological research. Conventional metrics such as Relative Bias (RB) and Absolute Relative Bias (ARB), although widely applied, exhibit several limitations. These include sensitivity to the scale of the population parameter, ambiguity regarding the direction of bias, and instability when the true parameter value is near zero. Such limitations can lead to misrepresentations of estimator performance. BTBA addresses these concerns through an integrated approach that combines standardized bias quantification, empirically informed decision criteria, and comprehensive visual diagnostics.

The first component of the BTBA framework is the $Z^*$ statistic, a simulation-specific standardized measure of bias. This metric centers each estimate on its known population value and scales deviations by the root mean squared error (RMSE), thereby enabling scale-invariant comparisons across different conditions. Unlike RB and ARB, $Z^*$ captures the distributional properties of bias by assessing how far estimates deviate from the true value relative to the expected variability. This transition from raw point-based measures to a standardized diagnostic enhances both the statistical rigor and interpretability of bias assessments.

The second component is the BTBA decision matrix, which classifies bias using the mean and variance of $Z^*$ into four interpretive categories: accept, accept with caution, research dependent and reject. These categories are defined based on empirically derived thresholds, informed by the distributional properties of the standard normal distribution in simulation related studies (Bollen, 1989; Mooney, 1997). By formalizing the criteria for bias evaluation, the decision matrix replaces subjective interpretation with a consistent and statistically grounded framework, thereby enhancing comparability across studies and conditions. One potential limitation involves the uniform application of $Z^*$ thresholds across varying parameter types and model contexts. Although the thresholds used in this study are informed by the standard normal distribution, future research may consider refining these benchmarks to account for natural differences in estimation precision across different parameter classes.

The third component involves the use of visual diagnostics, particularly ridgeline plots, to examine the distributional behavior of parameter estimates. These visualizations expose key characteristics of estimator performance, including skewness, variance inflation or deflation, multimodality, and proximity to the true value. By embedding descriptive statistics such as the mean and variance directly within the plots, BTBA enables a context-rich interpretation of $Z^*$ patterns grounded in the empirical behavior of the estimator.

The Bhirkuti's Test of Bias Acceptance (BTBA) framework offers a cohesive and multidimensional approach to evaluating bias in simulation research through the integration of standardized bias quantification, structured decision criteria, and visual diagnostics. It is particularly effective in complex modeling contexts such as latent growth curve models, small-sample designs, and planned missing data structures, where conventional metrics often fail to capture meaningful estimation patterns or may misrepresent estimator performance. Although the current application is situated within psychometric simulation, the BTBA framework holds substantial promise for broader use in domains where bias evaluation plays a critical role in simulation-based analysis (Carsey & Harden, 2013). Its methodological rigor, interpretability, and flexibility make it a valuable tool for assessing estimator behavior across diverse statistical models and research contexts (Enders, 2022; Mistler & Enders, 2012). By aligning simulation diagnostics with established inferential standards, BTBA represents a methodological advancement in bias evaluation. It provides a structured, transparent, and scalable framework that allows researchers to make more defensible and nuanced judgments about estimator quality. The integration of simulation-specific standardized metrics, empirically derived decision rules, and informative visual tools enhances both the robustness and interpretability of simulation findings, supporting more credible and consistent assessments across varied analytic conditions.

**Funding Details**

This research received no specific grant from any funding agency, commercial, or not-for-profit sectors.

**Disclosure Statement**

**Data Availability Statement**

The data and analysis scripts used in this study are available upon request. Please contact Aneel Bhusal at abhusal@ttu.edu for access.

**AI Disclosure**

AI has been used for editing, proofreading, and coding in the preparation of this manuscript.

**Additional Information**

*Analysis of additional parameters and additional information is available on request to Aneel Bhusal aneelbhusal@gmail.com.*